\documentclass[journal=jacsat,manuscript=article]{achemso}

\usepackage[version=4]{mhchem} % Formula subscripts using \ce{}
\usepackage{xcolor}
\usepackage{amsmath}
\usepackage{gensymb}

\usepackage{xr}

\makeatletter
\newcommand*{\addFileDependency}[1]{% argument=file name and extension
  \typeout{(#1)}
  \@addtofilelist{#1}
  \IfFileExists{#1}{}{\typeout{No file #1.}}
}
\makeatother

\newcommand*{\myexternaldocument}[1]{%
    \externaldocument{#1}%
    \addFileDependency{#1.tex}%
    \addFileDependency{#1.aux}%
}

\myexternaldocument{SupportingInformation}
\usepackage{etoolbox}% needed for the patch  <<<

\makeatletter
\renewcommand*{\acs@author@fnsymbol@symbol}[1]{% Use numbers instead of symbols, * is for email
    \ifcase #1 *\or
    1\or
    2\or
    a\or
    4\or
    5\or
    6\or
    7\or
    8\or
    9\or
    10
    \fi
}
        
\renewcommand*\acs@contact@details{% addd * before  E-mail
    {\sffamily *\,E-mail: \acs@email@list }%
    \acs@number@list
}           

\patchcmd{\acs@address@list@auxii}% superscript for numbers in affiliations
{\acs@author@fnsymbol{\acs@affil@marker@cnt}}
{\textsuperscript{\acs@author@fnsymbol{\acs@affil@marker@cnt}}}
{}{}

\patchcmd{\acs@address@list@auxii}% superscript for numbers in affiliations
{{\acs@author@fnsymbol{\acs@affil@marker@cnt}\@nameuse{@altaffil@\@roman\@tempcnta}\par}}
{{\textsuperscript{\acs@author@fnsymbol{\acs@affil@marker@cnt}}\@nameuse{@altaffil@\@roman\@tempcnta}\par}}
{}{}        

\makeatother    
\author{Christopher G. Bailey}
\affiliation[ACEx]{ARC Centre of Excellence in Exciton Science, School of Physics, University of New South Wales, Sydney, Australia}
\altaffiliation{Both authors contributed equally}
\email{Christopher.Bailey@unsw.edu.au}
\author{Lara V. Gillan}
\affiliation[ACEx]{ARC Centre of Excellence in Exciton Science, School of Physics, University of New South Wales, Sydney, Australia}
\altaffiliation{Both authors contributed equally}
\author{Minwoo Lee}
\affiliation[SPREE]
{School of Photovoltaics and Renewable Energy Engineering, University of New South Wales, Sydney, Australia}
\author{Nicholas Sloane}
\affiliation[ACEx]{ARC Centre of Excellence in Exciton Science, School of Physics, University of New South Wales, Sydney, Australia}
\author{Xu Liu}
\affiliation[SPREE]
{School of Photovoltaics and Renewable Energy Engineering, University of New South Wales, Sydney, Australia}
\author{Arman Mahboubi Soufiani}
\affiliation[SPREE]
{School of Photovoltaics and Renewable Energy Engineering, University of New South Wales, Sydney, Australia}
\author{Xiaojing Hao}
\affiliation[SPREE]
{School of Photovoltaics and Renewable Energy Engineering, University of New South Wales, Sydney, Australia}
\author{Dane R. McCamey}
\affiliation[ACEx]{ARC Centre of Excellence in Exciton Science, School of Physics, University of New South Wales, Sydney, Australia}
\email{Dane.McCamey@unsw.edu.au}

\title[Influence of Organic Spacer Cation on Dark Excitons in 2D Perovskites]
  {Influence of Organic Spacer Cation on Dark Excitons in 2D Perovskites}

\begin{document}


% \begin{tocentry}

% Some journals require a graphical entry for the Table of Contents.
% This should be laid out ``print ready'' so that the sizing of the

% Inside the \texttt{tocentry} environment, the font used is Helvetica
% 8\,pt, as required by \emph{Journal of the American Chemical

% The surrounding frame is 9\,cm by 3.5\,cm, which is the maximum
% permitted for  \emph{Journal of the American Chemical Society}
% graphical table of content entries. The box will not resize if the
% content is too big: instead it will overflow the edge of the box.

% This box and the associated title will always be printed on a
% separate page at the end of the document.

% \end{tocentry}

\begin{abstract}
  The organic spacer cation plays a crucial role in determining the exciton fine structure in two-dimensional (2D) perovskites. Here, we use low-temperature magneto-optical spectroscopy to gain insight into the influence of the organic spacer on dark excitons in Ruddlesden--Popper (RP) perovskites. We show that by using modest magnetic field strengths ($<$1.5 T), the spin-forbidden dark-exciton state can be identified and its emission properties significantly modulated through the application of in-plane magnetic fields, up to temperatures of 15 K. At low temperatures, an increase in collected photoluminescence efficiency of  $>$30\% is demonstrated, signifying the critical role of the dark exciton state for light-emitting applications of 2D perovskites. The exciton fine structure and the degree of magnetic-field-induced mixing are significantly impacted by the choice of organic spacer cation, with 4‐methoxyphenylethylammonium (MeO-PEA) showing the largest effect due to larger bright--dark exciton splitting. Our results suggest that dark excitons preferentially form biexcitons depending on the choice of spacer. We distinguish between interior (bulk) and surface dark-exciton emission, showing that bright--dark exciton splitting differs between the interior and surface. Our results emphasise the significance of the organic spacer cation in controlling the exciton fine structure in 2D perovskites and have important implications for the development of optoelectronic technology based on 2D perovskites.
\end{abstract}

\section{Introduction}
Two-dimensional (2D) perovskites have attracted significant attention in recent years due to their unique optoelectronic and structural properties. These materials consist of inorganic perovskite layers separated by organic cations which provide criteria that violate the Goldschmidt tolerance factor, forming natural quantum wells of varying thicknesses. \cite{Ishihara1989,Papavassiliou1993,Ishihara1990OpticalCnH2n+1NH32PbI4,Hong1992} Unlike their three-dimensional (3D) counterparts, 2D perovskites exhibit strong excitonic effects due to quantum confinement and reduced dielectric screening caused by the mismatch in dielectric constant between the inorganic sheets and organic spacers \cite{Kim2015MulticoloredDiodes,Ishihara1989,Papavassiliou1993,Ishihara1990OpticalCnH2n+1NH32PbI4,Hong1992}.

2D perovskites have shown promise for optoelectronic applications, including solar cells, \cite{Zhao2022,Zhao2022a,ThrithamarasseryGangadharan2019,Leung2022,Chen2018} light-emitting diodes, \cite{Shang2019,Yuan2019,Pang2020,Worku2021,Yang2018} and lasers. \cite{Wang2022,Zhang2018,Qin2020} The incorporation of 2D perovskites has led to the demonstration of solar cells with enhanced stability,\cite{Zhao2022,Zhao2022a,ThrithamarasseryGangadharan2019,Leung2022,Chen2018} including solar cells with an operational lifetime exceeding one year. \cite{Grancini2017} However, a comprehensive understanding of their electronic and optical properties is currently lacking, despite being crucial for further advancements in these technologies.

The increased quantum and dielectric confinement produced by the structure of 2D perovskites leads to a large exciton binding energy ($E_{B}$) in the range of 150--750 meV. \cite{Hong1992,Koutselas1996ElectronicUnits, Takagi2013InfluenceMaterial, Mousdis2000PreparationStructure, Dammak2009Two-dimensionalNanomaterial,Ishihara1990OpticalCnH2n+1NH32PbI4,Liao2022Inorganic-CationCrystal}
These values of $E_{B}$ are substantially greater than those of bulk 3D perovskites and even 3D-perovskite nanocrystals, resulting in the dominance of excitons as the photoexcited species at room temperature. They exhibit a rich exciton fine structure due to mechanisms such as the electron--hole exchange interaction and crystal field splitting, which remove the degeneracy of the exciton states, allowing them to be discerned via methods such as optical spectroscopy \cite{Dyksik2021BrighteningPerovskites,Posmyk2022QuantificationCompound,Tang2021ThicknessField,Canet-Albiach2022RevealingPassivation}. 

Ruddlesden--Popper (RP) phase 2D perovskites have the general formula \ce{\textit{A}_{n+1}\textit{B}_{n}\textit{X}_{3n+1}}, where $A$ is often an organic cation, $B$ is a transition metal and $X$ is a halide anion. The most commonly studied composition is phenylethylammonium (phenethylammonium, PEA) lead iodide ($\mathrm{(PEA)_{2} PbI_{4}}$), characterised by single layers (n=1) of lead iodide octahedra separated by interlayers comprised of the bulky $\mathrm{PEA}$ cation. The choice of organic cation spacer has been demonstrated to affect dielectric environment, distortion angles of perovskite octahedral units, bandgap energy, effective mass of carriers, and exciton--phonon coupling \cite{Liu2019InfluencePerovskites,Paritmongkol2020TwoPerovskites,Koegel2022CorrelatingPerovskites}. Recently, functionalisation of the PEA cation has been investigated, where for example, a methoxy group is used to produce 4‐methoxyphenylethylammonium (MeO-PEA). MeO-PEA cations have been utilised as an additive for 3D-perovskite based solar cells to improve passivation \cite{Min2021,Meng2022} and also in quasi-2D perovskite solar cells for improved ambient stability. \cite{Fu2019} 

\begin{figure*}[htbp]
  \centering
  \includegraphics[width=0.85\textwidth]{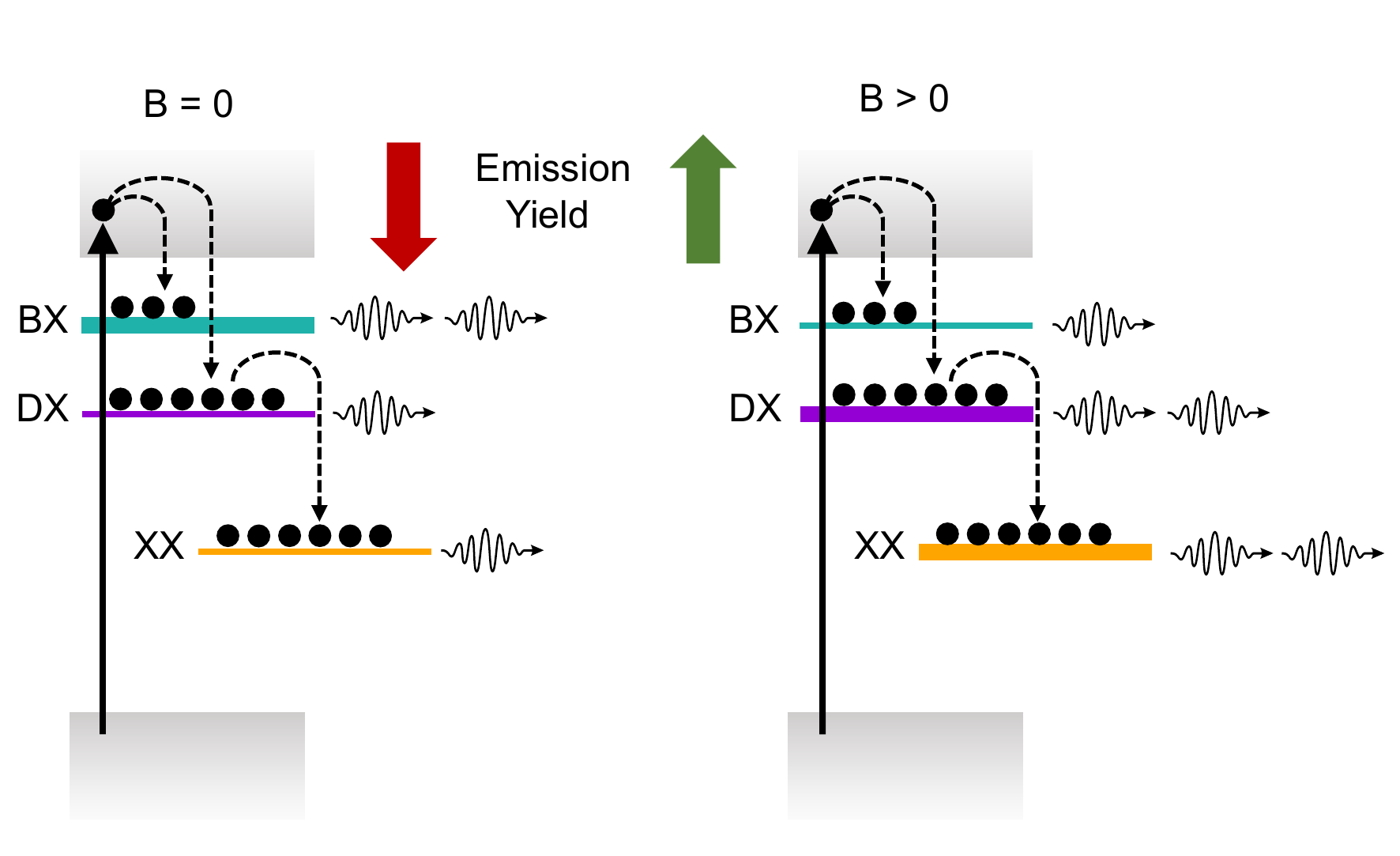}
  \caption{\textbf{Mechanism of magnetic-field-induced photoluminescence brightening at low temperature.} Following non-resonant photoexcitation, the dark exciton state, which also can form biexcitons, is preferentially populated at low temperatures. When an in-plane magnetic field is applied, oscillator strength is transferred from bright exciton states to the more highly populated dark state (represented by the line thickness) and the overall radiative emission intensity is increased.}
  \label{fig:mainschem}
\end{figure*}

Several studies have utilised magneto-optical spectroscopy to gain insight into the nature of the exciton fine structure of 2D perovskites, often using large magnetic fields ($>$ 50 T) to selectively manipulate exciton states according to their dipole orientation, hence modifying their optical characteristics \cite{Dyksik2021BrighteningPerovskites,Posmyk2022QuantificationCompound,Surrente2021PerspectiveField,Neumann2021ManganesePerovskites,Dyksik2021TuningConfinement,Tang2021ThicknessField,Do2020BrightPerovskites,Kahmann2021,Urban2020RevealingPerovskites}. Here, we show that applying a comparatively modest in-plane magnetic field ($<$ 1.5 T) significantly increases the photoluminescence intensity of 2D perovskite thin films in the family of \ce{\textit{A}2PbI4} at low temperatures (Figure \ref{fig:mainschem}), where the choice of cation, \textit{A}, has a substantial impact on this effect. We use these observations to gain insight into the exciton fine structure of 2D perovskites, revealing the role and behaviour of bright exciton, dark exciton and biexciton energy states in relation to the spacer cation used.

%Dyksik \textit{et al.} report the exciton fine structure in terms of the alignment of the oscillator electric dipole relative to the stacking direction of perovskite layers. By studying the polarisation dependence of transmittance for the zero-field and high field case they were able to assign the relative exciton \cite{dyksik2021brighteningperovskites}. The exchange interaction and inherent crystal field lift the degeneracy of the different exciton spin configurations such that the transitions become non-degenerate. Due to the, previously discussed, Rashba effect, two of the optically bright states are couple to circularly polarised light, these have been ascribed a electric dipole which is in-plane \cite{dyksik2021brighteningperovskites}. While a third high-energy bright state is found to have an out of plane dipole moment \cite{dyksik2021brighteningperovskites}. Upon the application of a magnetic field in the Voigt configuration the exciton sublevels split with the two in-plane sublevels mixing with the dark and out of plane exciton sublevels, such that the dark state and its mixed in-plane state couple to light with polarisation $E$ parallel to $B$, and the remaining two exciton states couple to polarised light $E\perp B$ \cite{dyksik2021brighteningperovskites}. This field dependent brightening of the dark excitonic state is of particular interest due to the long spin lifetime, which loans itself well to potential application in spintronic and quantum computing applications \cite{tang2021thickness}.

\section{Results and discussion}

\begin{figure*}[htbp]
  \centering
  \includegraphics[width=0.8\textwidth]{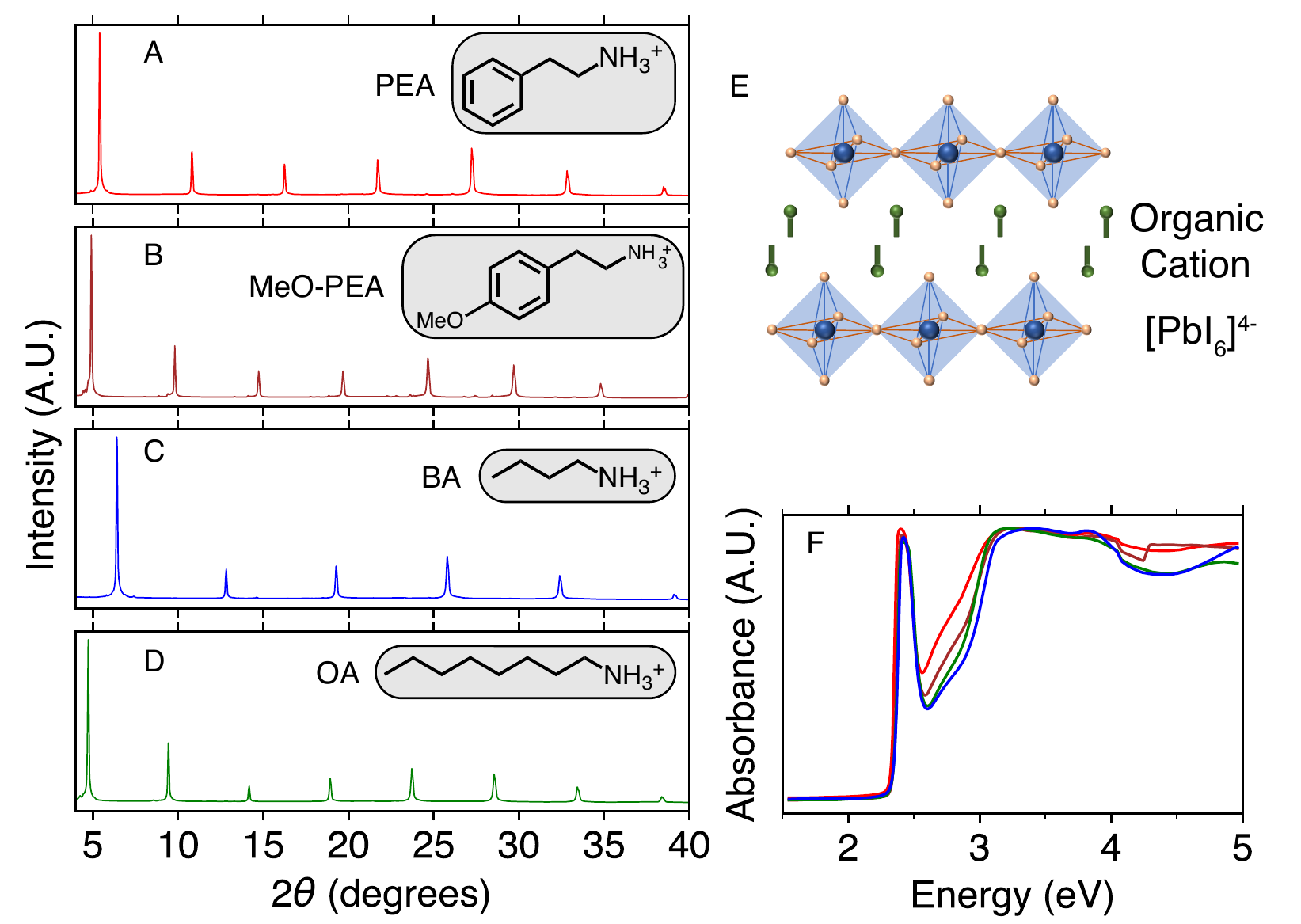}
  \caption{\textbf{Characterisation of the 2D perovskite thin films.} X-ray diffractograms for (A) \ce{(PEA)2PbI4}, (B) \ce{(MeO-PEA)2PbI4}, (C) \ce{(BA)2PbI4}, (D) \ce{(OA)2PbI4} showing typical diffraction patterns for 2D perovskite thin films. (E) Structure of 2D perovskite crystal of alternating organic cations and inorganic \ce{[PbI6]^4-} with n=1. (F) UV-vis absorbance spectra for \ce{(PEA)2PbI4} (red), \ce{(MeO-PEA)2PbI4} (brown), \ce{(BA)2PbI4} (blue), and \ce{(OA)2PbI4} (green), showing sharp exciton-like transitions.}
  \label{fig:char}
\end{figure*}

We firstly produced the following RP phase perovskites: \ce{(PEA)2PbI4}, \ce{(MeO-PEA)2PbI4}, \ce{(BA)2PbI4}, and \ce{(OA)2PbI4}, where PEA is phenylethylammonium, MeO-PEA is PEA with a methoxy functional group (4‐methoxyphenylethylammonium), BA is \textit{n}-butylammonium, and OA is \textit{n}-octylammonium. Films with a thickness of $\sim$350-400 nm were obtained by spin coating the precursor solutions onto quartz substrates using the procedure described in the experimental section. Characterisation results of the 2D perovskite thin films are shown in Figure \ref{fig:char}. We performed X-ray diffraction (XRD) on the samples (Figure \ref{fig:char}A--D), with results showing typical patterns for the 2D (n=1) RP phase perovskite structure shown in Figure \ref{fig:char}E. UV-visible absorption spectroscopy (room temperature) shows that the samples have a sharp optical transition around 2.4 eV, indicative of materials with a large exciton binding energy (Figure \ref{fig:char}F).

%\textcolor{red}{The larger exciton binding energy observed in conjugated phenylethylammonium layers is attributed to its higher dielectric constant compared to saturated butylammonium molecular layers.\cite{Yaffe2015ExcitonsCrystals,Dammak2009Two-dimensionalNanomaterial}}

\begin{figure*}[htbp]
  \centering
  \includegraphics[width=0.95\textwidth]{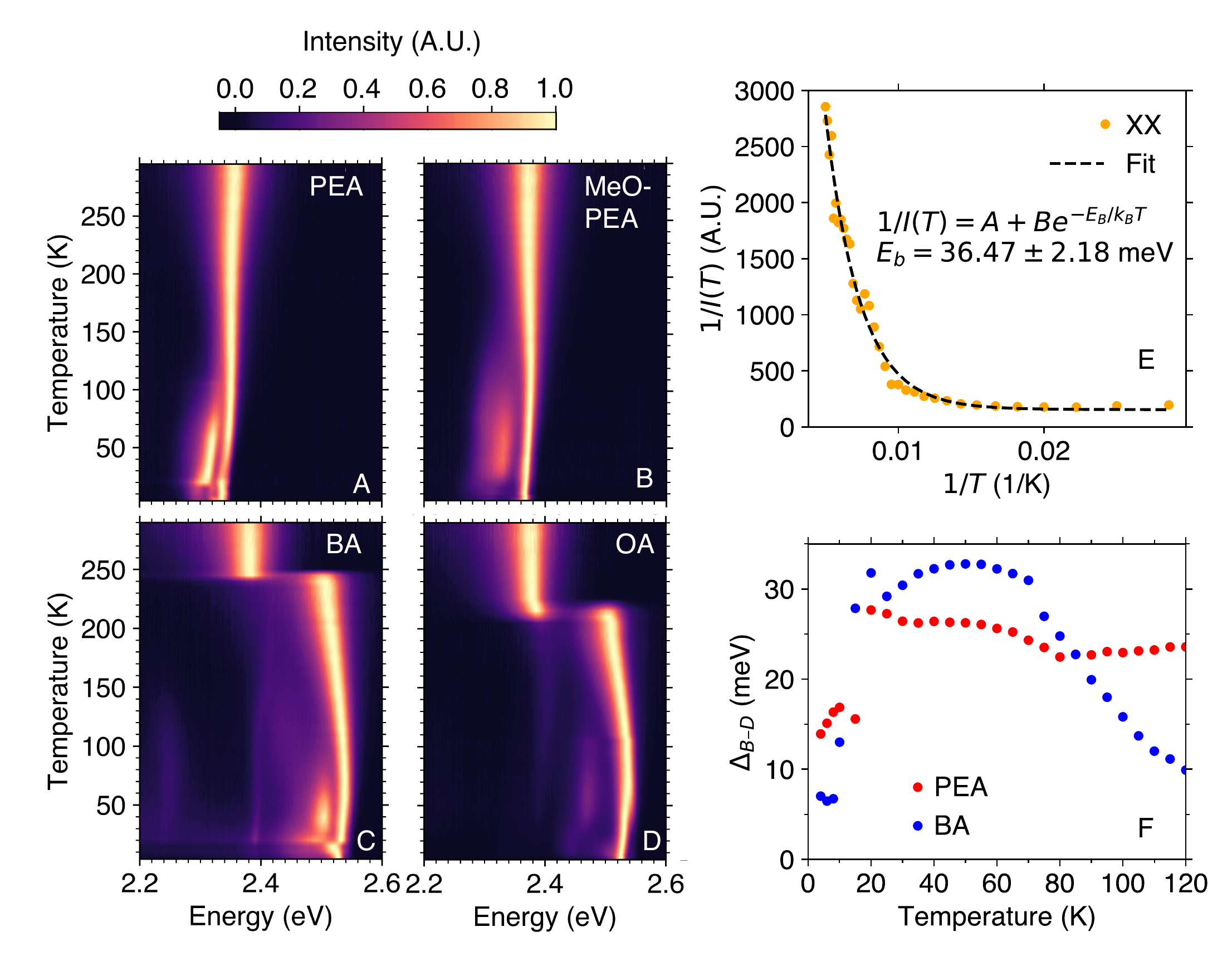}
  \caption{\textbf{Temperature-dependent photoluminescence.} False-colour map of normalised photoluminescence spectra for (A) \ce{(PEA)2PbI4}, (C) \ce{(MeO-PEA)2PbI4}, (B) \ce{(BA)2PbI4}, and (D) \ce{(OA)2PbI4}. (E) Inverse integrated intensity ($(1/I(T)$) of the biexciton emission versus inverse temperature ($1/T$) for \ce{(PEA)2PbI4}, fitted with equation \ref{eqn:E_B}. The biexciton binding energy is estimated from the activation energy for dissociation. F) Bright--dark exciton splitting, $\Delta_{B-D}$ as a function of temperature, comparing \ce{(PEA)2PbI4} and \ce{(BA)2PbI4}.}
  \label{fig:Tempdep}
\end{figure*}

To investigate the nature of the exciton fine structure in 2D perovskite thin films we first carried out measurements of steady-state photoluminescence in the temperature range of 2.4--300 K, as shown in Figure \ref{fig:Tempdep}. At room temperature, all samples exhibit a broad emission with a linewidth (full-width at half maximum) of  $\sim$1 eV, with a slightly different peak emission energy for each spacer cation. Upon cooling the sample from room temperature, the emission for all compositions undergoes a slight redshift, however, the most notable feature is an abrupt shift of the emission energy at $\sim$245 K in \ce{(BA)2PbI4} and at $\sim$210-215 K in \ce{(OA)2PbI4}. In general, such features are often observed at temperatures where structural phase changes occur \cite{Piana2019,Kong2015a}, and in the 2D RP phase perovskites studied, this has been attributed to a change in relative alignment of \ce{PbI4} octahedra, caused by ordering of weakly interacting alyklammonium chains \cite{Ziegler2022,Menahem2021,Barman2003,Ishihara1990OpticalCnH2n+1NH32PbI4,Yaffe2015ExcitonsCrystals}. This is not observed in \ce{(PEA)2PbI4} or \ce{(MeO-PEA)2PbI4}, due to the phenylethylammonium chains being shorter and less flexible than \textit{n}-butylammonium and \textit{n}-octylammonium. \cite{Menahem2021,Ishihara1990OpticalCnH2n+1NH32PbI4} 

On further decreasing temperature below $\sim$200 K, the \ce{(BA)2PbI4} and \ce{(OA)2PbI4} samples exhibit an emission band between 2.38--2.42 eV, which undergoes a slight redshift as the temperature is decreased. In the \ce{(BA)2PbI4} thin film, this emission is narrower and persists down to low temperatures, whereas in \ce{(OA)2PbI4} the emission is undetectable below $\sim$40 K. In \ce{(OA)2PbI4}, it is likely that the origin of this emission is due to a persistence of the higher-temperature phase with distorted \ce{PbI4} octahedra, since the lineshape is consistent at temperatures above and below the transition temperature of the phase change. On cooling further, the spectrum eventually fully transitions to that of the low temperature phase.

For the low-energy emission in \ce{(BA)2PbI4}, similar features have been observed previously and were attributed to emission from the interior, in contrast to the surface emission which was assigned to the high energy peak. \cite{Wang2023Exciton-phononMicroplates,Sheikh2018PossibleFew-Layer} Several other studies report origins such as magnetic dipole emission from self-trapped excitons, \cite{Decrescent2020Even-ParityPerovskites} the effect of strain induced by cation stacking,\cite{Du2020StackingManipulation} and interactions of the lead iodide interlayers.\cite{Sheikh2020MolecularPerovskites} Regardless of the physical mechanism responsible, it is clear from the previous literature that the emission source can be separated into interior and surface emission. \cite{Wang2023Exciton-phononMicroplates,Sheikh2018PossibleFew-Layer}

For \ce{(PEA)2PbI4} at low temperatures, three optical transitions can be easily distinguished, which themselves are known to be multi-component (Figure S2). \cite{Canet-Albiach2022RevealingPassivation,Dyksik2021BrighteningPerovskites, Posmyk2022QuantificationCompound,Tang2021ThicknessField}. Four primary exciton states have previously been identified: a 'dark' exciton state in the singlet spin configuration with zero angular momentum ($J=0$) and three optically active ('bright') states with $J=1$, which in general are split by exchange interaction. In 2D perovskites, these bright states are typically split into states with in-plane and out-of-plane dipole moments due to their alternating layered structure having a broken symmetry perpendicular to the axis of the stacking \cite{Dyksik2021BrighteningPerovskites,Tamarat2019TheState,Hou2021RevealingNanocrystals,Posmyk2022QuantificationCompound}. 

To avoid potential error from fitting a non-unique set of transitions, we separate the observed emission in to three main components: we attribute the highest energy peak to emission from the bright exciton transitions (BX), the intermediate energy peak to the spin-forbidden dark exciton transition (DX), and the lowest energy peak to a biexciton transition (XX) (Figure S2).\cite{Tang2021ThicknessField,Dyksik2021BrighteningPerovskites,Posmyk2022QuantificationCompound,Thouin2018StableDisorder,Fang2020Band-EdgePerovskites,Li2020BiexcitonsCrystals}. We fitted the sum of three Gaussian functions (see supporting information) to the data obtained for each composition and found that these fit the data well over the temperature range 4--200 K, with additional transitions present in \ce{(BA)2PbI4} and \ce{(OA)2PbI4} as discussed in more detail below (Figures S2--S4). For \ce{(MeO-PEA)2PbI4}, the addition of a low-energy tail overlapping with the XX transition (Figures \ref{fig:Tempdep}B and S5) means that the spectrum does not fit well with three, or even four Gaussian transitions due to the asymmetry introduced. This suggests that \ce{(MeO-PEA)2PbI4} has additional disorder or a defect state which is not present in \ce{(PEA)2PbI4}, but may share the same origin of the lower energy shoulder in the BA and OA compositions (Figures S3 and S4).

A biexciton is the simplest type of multiexciton state, formed from two excitons and typically observed at higher excitation intensities. These states have an energy which is lower than the single-exciton state, separated by the biexciton binding energy. Relaxation to the singly excited state can occur through the photon emission (radiative) or via non-radiative Auger processes. The presence of photoexcited biexcitons in 2D RP phase perovskites has been the subject of debate \cite{Kahmann2021,Thouin2018StableDisorder,Fang2020Band-EdgePerovskites,Li2020BiexcitonsCrystals} but despite this, evidence has been produced that suggests biexcitons in \ce{(PEA)2PbI4} persist at room temperature \cite{Thouin2018StableDisorder}, and the biexciton binding energy has been previously determined to be 39-49 meV. \cite{Thouin2018StableDisorder,Fang2020Band-EdgePerovskites,Li2020BiexcitonsCrystals} By extracting and fitting the temperature-dependent intensity of the biexciton emission, we can estimate the biexciton binding energy as an 'activation energy' for dissociation by using the following exponential equation \cite{Li2016c,Zhang2020,Liao2022Inorganic-CationCrystal} (see supporting information for details), 

\begin{equation}
  1/I(T) = A + B e^\frac{-E_B}{Tk_B}
  \label{eqn:E_B},
\end{equation}

 \noindent where $A$ is the inverse of the photoluminescence intensity at 0 K, $(1/I(0)$), and $B$ is the ratio of the biexciton dissociation rate to the emission rate, $k_\text{dis}/k_r$.
 
We applied this fitting process to the data in the temperature range of 30--200 K for \ce{(PEA)2PbI4}, where the intensity decreases monotonically and smoothly (Figure S1--S3). The result of this fitting is shown in Figure \ref{fig:Tempdep}E, and a value of 36.47±2.18 meV is obtained for $E_B$, which is slightly lower than the range of 39-49 meV determined in prior work \cite{Thouin2018StableDisorder,Fang2020Band-EdgePerovskites}. We note that this method has several limitations: for example, it does not account for a binding energy which varies with temperature (ignoring possible phase changes) and assumes that the phonon-assisted dissociation is the dominant mechanism competing with emission when increasing temperature. A very similar value of activation energy (36.7 meV) was previously determined in \ce{(PEA)2SnI4} and attributed to the activation energy for thermal re-excitation from lower energy states, proposed to originate from bound excitons or shallow defects \cite{Kahmann2021}. This prior work could therefore indicate a similar biexciton binding energy in \ce{(PEA)2SnI4}, but we do not completely rule out the possibility of the XX-assigned transition being due to emission from such singly-excited states.

To provide further insight into the photophysics of the exciton states discussed, we performed photoluminescence measurements at low temperatures with a magnetic field applied parallel to the plane of the substrate and interlayers, perpendicular to the direction of the excitation beam (Voigt configuration), as described in the experimental section. 

\begin{figure*}[htbp]
  \centering
  \includegraphics[width=0.95\textwidth]{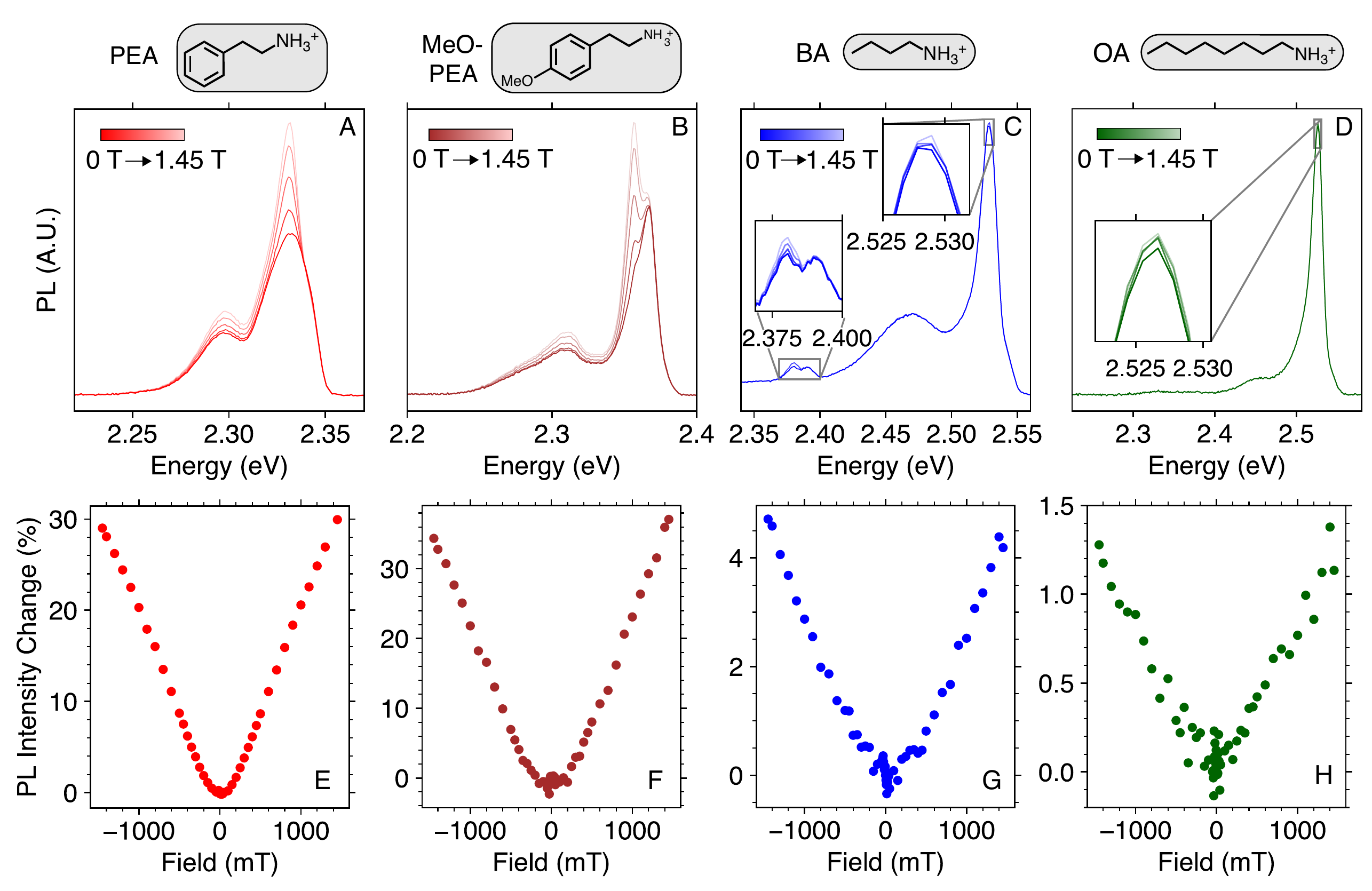}
  \caption{\textbf{Photoluminescence of 2D perovskite thin films at 2.4 K in the presence of an in-plane magnetic field.} Low temperature, magnetic-field dependent photoluminescence spectra for (A) \ce{(PEA)2PbI4}, (B) \ce{(MeO-PEA)2PbI4}, (C) \ce{(BA)2PbI4}, (D) \ce{(OA)2PbI4}. The corresponding change in photoluminescence intensity across regions where change is detected for (E) \ce{(PEA)2PbI4} (2.27--2.35 eV), (F) \ce{(MeO-PEA)2PbI4} (2.27--2.4 eV) (G), \ce{(BA)2PbI4} (2.37--2.40 eV and 2.50--2.53), and (H) \ce{(OA)2PbI4} (2.50--2.53 eV).}
  \label{fig:MPL}
\end{figure*}

Figure \ref{fig:MPL} shows the photoluminescence spectrum as a function of magnetic field at a temperature of 2.4 K for the different spacer cations used. At this temperature, the total integrated intensity increases monotonically with increasing magnetic field for all compositions. This can be almost directly translated to an increase in photoluminescence quantum yield (PLQY) within the collection scheme used, since the change in absorption under applied fields of $<$ 1.5 T has been previously found to be negligible \cite{Dyksik2021BrighteningPerovskites}.

This effect is much more pronounced in PEA and MeO-PEA, with $>$ 30\% enhancement of integrated total photoluminescence intensity (Figures \ref{fig:MPL}A, B, E and F). PEA-based 2D perovskites have been the most successfully utilised compositions for optoelectronic devices such as solar cells and LEDs,\cite{Yang2018,ThrithamarasseryGangadharan2019,Zhao2022a,Leung2022} which alludes to a potential link between observed dark-exciton brightening and performance of materials. For the other cations, BA (Figure \ref{fig:MPL}C and G) and OA (Figure \ref{fig:MPL}D and H), the detectable total photoluminescence enhancement is limited to a very narrow region of the emission and overall is very small ($<<$ 1\%). Nonetheless, the modulation of the photoluminescence is significant enough to be identified using the simple detection method used (see experimental section) and can be spectrally resolved, offering important insights into the origin of the emission. 

Dyksik \textit{et al.} describe the magnetic field effects observed in 2D RP perovskites as mixing of dark and bright states caused by the in-plane magnetic field breaking the crystal field symmetry, which results in a transfer of oscillator strength and population \cite{Dyksik2021BrighteningPerovskites}. At low temperatures, the lower-energy dark exciton state has a higher probability of occupation than the bright exciton states following non-resonant photoexcitation and subsequent thermal relaxation. Therefore, a transfer of oscillator strength to the dark exciton state would increase the total photoluminescence quantum yield if the population difference is large enough, despite the dark-exciton radiative emission being spin-forbidden. This should result in a decrease of magnetic-field-induced photoluminescence brightening with increasing temperature, which explains why the magnetic field effect was not detectable above 2.4 K in \ce{(BA)2PbI4} and \ce{(OA)2PbI4}, and the total enhancement decreases with temperature up to 15 K in \ce{(PEA)2PbI4} and \ce{(MeO-PEA)2PbI4}, beyond which the effect was found to diminish (Figures S7 and S8). At higher temperatures, the reduction in oscillator strength of the bright exciton out-competes the dark-exciton brightening at low fields due to the smaller difference in population at these temperatures. Hence, the integrated photoluminescence is actually reduced between 0--0.5 T at temperatures 10 and 15 K, beyond which it starts to increase again to $>$ +10\% at 1.45 T (Figure S9). 

An increase in photoluminescence in \ce{(BA)2PbI4} and \ce{(OA)2PbI4} is observed in the region of the highest intensity peak, attributed to bright-exciton emission (Figures \ref{fig:MPL}C and D). This increase is likely indicative of lower bright--dark exciton splitting in these compositions (Figure \ref{fig:Tempdep}F), rendering these states too close in energy to distinguish in terms of magnetic-field brightening. Fitting the low-temperature photoluminescence spectra to multiple contributions exemplifies this, with the main peak of the data fitting well to two transitions (bright and dark excitons) closer together in energy for \ce{(BA)2PbI4} and \ce{(OA)2PbI4} (Figure S3 and S4). From this analysis, the bright--dark splitting is estimated to be $\sim$7 meV for \ce{(OA)2PbI4} and \ce{(BA)2PbI4}, compared to $\sim$12--14 meV for \ce{(PEA)2PbI4} and \ce{(MeO-PEA)2PbI4}. This corroborates with a reduced magnetic-field effect in \ce{(OA)2PbI4} and \ce{(BA)2PbI4} compared to \ce{(PEA)2PbI4}, since the energy difference between the bright and dark exciton states is half the size, and hence the population difference is smaller. 

\begin{figure*}[htbp]
  \centering
  \includegraphics[width=0.75\textwidth]{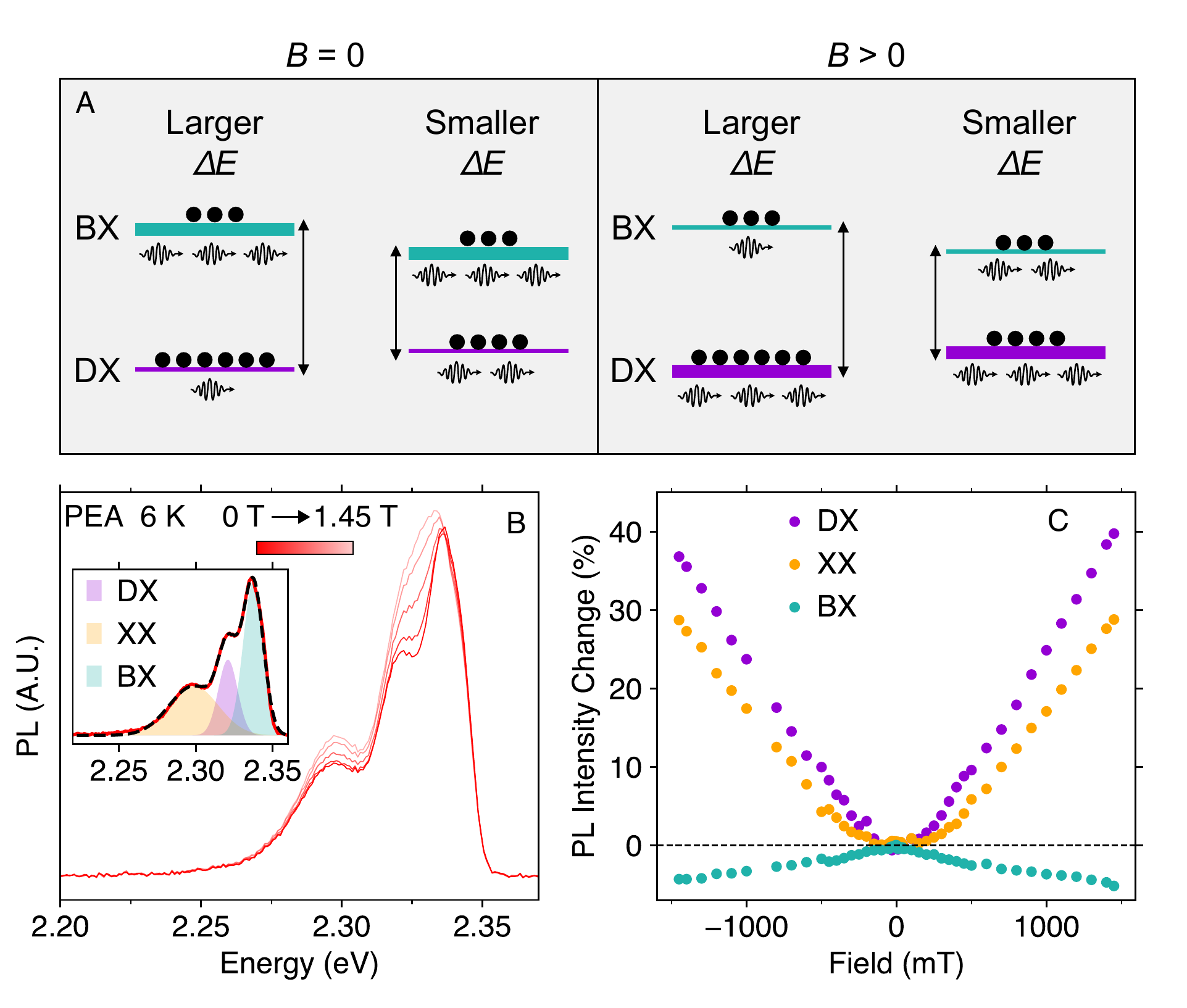}
  \caption{\textbf{Response of individual exciton transitions to an applied in-plane magnetic field.}  (A) Schematic showing the effect of bright--dark exciton splitting on the magnetic-field-induced brightening. The width of the lines represent the oscillator strength of the transitions. For larger bright--dark exciton splitting, the population difference is larger and when the in-plane magnetic field is applied, a large difference in photoluminescence intensity is observed when oscillator strength is transferred. For a smaller splitting, the population difference is smaller and hence when the field is applied, the brightening is less pronounced. (B) Magnetic-field dependent photoluminescence spectra of \ce{(PEA)2PbI4} at 6 K. The inset shows the spectral decomposition by fitting multiple Gaussian functions. (C) Integrated intensity (amplitude) of Gaussian components versus magnetic field for the individual bright exciton (BX), dark exciton (DX), and biexciton (XX) transitions.}
  \label{fig:MPL_PEA}
\end{figure*}

Figure \ref{fig:MPL_PEA}A illustrates this relationship between the observed dark-exciton brightening and the size of the bright--dark energy splitting. The difference in population is likely to depart significantly from one based purely on Boltzmann-like statistics, since non-thermalised ('hot') photoluminescence is known to occur in these materials, reducing the proportion of excitons that fully relax to the dark exciton state following non-resonant excitation. \cite{Kahmann2021,Dyksik2021BrighteningPerovskites} 

The diamagnetic coefficient in 2D perovskites containing an aromatic spacer, such as PEA, is larger compared to layered perovskites of the same thickness containing an aliphatic spacer. This is because the exciton wavefunction in layered perovskites with an aromatic spacer extends over a larger spatial area due to a smaller dielectric mismatch between the well and the barrier, compared to 2D perovskites with aliphatic spacers. \cite{Hong1992,Surrente2021PerspectiveField} This should result in a larger exchange interaction between the electron and hole, increasing the bright--dark exciton splitting in \ce{(PEA)2PbI4} and \ce{(MeO-PEA)2PbI4} compared with the other two compositions, in agreement with our observations. 

In the temperature-dependent photoluminescence data, (Figure \ref{fig:Tempdep}A--C and Figures S2--S3) the dark and bright exciton states seem to abruptly split further on increasing temperature beyond 20 K, with the apparent dark exciton state shifting to a lower energy, while the bright exciton emission energy is almost unchanged. This is accompanied by a concurrent increase in intensity of both the dark and biexciton states. We find that these results are repeatable and have been observed in previous studies, with the process responsible for the abrupt energy shift remaining unclear \cite{Kahmann2021}. The correlated increase in intensity with energy shift can be again explained via Boltzmann statistics, where the reduction in energy of the dark exciton results in a larger population, promoting radiative emission. As the temperature is further increased, the relative intensity from the biexciton and dark exciton decreases, as expected when thermal energy is supplied to the system. Interestingly, the shift at 20 K coincides with the temperature above which no magnetic field effects can be observed, suggesting that the low-temperature regime plays a crucial role in the dark-exciton brightening. 

The photoluminescence spectrum of \ce{(BA)2PbI4} at low temperatures exhibits two additional peaks at lower energy, which are not observed in the spectra of the other compositions (Figure \ref{fig:Tempdep}C and \ref{fig:MPL}C). As discussed above, it is possible that the emission is from the interior of crystal domains. \cite{Wang2023Exciton-phononMicroplates,Sheikh2018PossibleFew-Layer}  In this case, the two lower energy peaks observed likely originate from dark and bright exciton states of the interior separated by $\sim$10 meV, where the lower energy dark-exciton emission is brightened by the in-plane magnetic field. The bright--dark exciton splitting of the interior domains is larger than that of the surface, such that the brightening of the dark state can be distinguished, unlike the emission from the surface due to its lower bright--dark exciton splitting. 

%Additionally, the use of aromatic spacers causes less structural distortion of the inorganic sub-lattice, which leads to smaller effective masses in 2D perovskites containing an aromatic spacer compared to those with aliphatic spacers. \cite{Dyksik2020BroadPerovskites,Dyksik2021TuningConfinement,Surrente2021PerspectiveField}.} 

As mentioned, above temperatures of 2.4 K, the photoluminescence change with magnetic field was undetectable in \ce{(BA)2PbI4} and \ce{(OA)2PbI4}. In \ce{(PEA)2PbI4} and \ce{(MeO-PEA)2PbI4} however, the change in photoluminescence with magnetic field can be observed up to temperatures of 15 K (Figures S7 and S8). Figure \ref{fig:MPL_PEA} shows the magnetic-field-dependent photoluminescence for \ce{(PEA)2PbI4} at 6 K. At this temperature we were able to resolve the transitions more clearly due to the larger bright--dark splitting compared with 2.4 K, which aids with our confidence in the multi-component fitting (supporting information). We show how the integrated intensity of each component changes as a function of magnetic field by again fitting the sum of three Gaussian functions to the photoluminescence spectrum (Figure \ref{fig:MPL_PEA}B). At each temperature, we first fit the zero-field data as a reference to obtain initial parameters, then fix the widths of the Gaussian components before fitting the field-dependent data, setting the amplitude and position as free parameters. As the applied magnetic field increases, the dark-exciton and biexciton emission intensities increase, whereas the bright-exciton emission decreases (Figure \ref{fig:MPL_PEA}C), in agreement with previous reports. \cite{Dyksik2021BrighteningPerovskites, Posmyk2022QuantificationCompound,Tang2021ThicknessField}

  In \ce{CsPbI3} nanocrystals, previous photon correlation measurements indicate that the dark-exciton ground state favours the creation of biexcitons at low temperatures \cite{Tamarat2020TheNanocrystals}. If this is also the case for \ce{(PEA)2PbI4}, it would explain the correlation of the magnetic-field brightening observed for the dark exciton and biexciton transitions at 6 K (Figure \ref{fig:MPL_PEA}C) and it is less likely that only dark excitons form bound states or are preferentially trapped by shallow defects. These factors, combined with the simultaneous increase in emission of the two transitions at $\sim$ 20 K discussed above provides compelling evidence that biexcitons are formed from dark excitons in \ce{(PEA)2PbI4} and \ce{(MeO-PEA)2PbI4}, as they are in \ce{CsPbI3} nanocrystals. 

  At first glance, this interpretation appears to contradict the centre-energy positions of the emission peaks observed, since the difference in energy between the biexciton and dark exciton emission ranges from $\sim$25 meV at 4 K, to $\sim$18 meV at 120 K (Figure S2). However, it is difficult to estimate the biexciton binding energy directly from the positions of emission peaks since the small Stokes shift and the overlap of different spectral contributions from the exciton fine structure cause the precise energetic position of the exciton emission to become contaminated by self-absorption effects, leading to a systematic underestimation of the binding energy ($E_{B}$). This explains why the difference in energy between the dark and biexciton emission peaks is significantly lower than the value of $\sim$ 36.5 meV for the biexciton binding energy obtained from equation \ref{eqn:E_B}. Moreover, these considerations oppose the possibility of biexcitons forming from (or dissociating into) one of the bright exciton states, since the spectral energy difference between these is larger than 36.5 meV, instead ranging from $\sim$40-42 meV.

\newpage

\section{Conclusion}

In conclusion, our study reveals the crucial role of organic spacer cations in determining the exciton fine structure in two-dimensional perovskites. By using modest magnetic fields, we have been able to identify and modulate the emission properties of the optically inactive dark exciton state, which is known to play a vital role in semiconductor light emission processes. The degree of magnetic-field-induced mixing and the characteristics of these states were found to be significantly influenced by the choice of spacer cation, with phenylethylammonium and 4-methoxyphenylethylammonium exhibiting the largest effect. Since PEA-based compositions have been the most successfully utilised materials for 2D perovskite optoelectronics in the field, our work highlights a potential correlation between observed dark-exciton brightening and material performance. We have found significant evidence that biexcitons are formed from dark excitons in \ce{(PEA)2PbI4} and \ce{(MeO-PEA)2PbI4} and mimic their magnetic-field-induced photoluminescence brightening. In \ce{(BA)2PbI4}, the surface and interior fine structure can be distinguished, with different bright--dark exciton splitting between the two emission sources. These findings provide valuable insights into the underlying mechanisms behind the exciton fine structure in 2D perovskites and highlight the importance of the organic spacer cation in controlling their optoelectronic properties. Our results are important for the design and development of next-generation, efficient light-emitting devices and other optoelectronic technologies based on 2D perovskites. 

\section{Experimental Section}

\textit{Materials}: Lead iodide (\ce{PbI2}, 99.999\%) was purchased from TCI. N,N-dimethylformamide (DMF, 99.99\%) and dimethyl sulfoxide (DMSO, 99.50\%) were purchased from Sigma-Aldrich. Phenethylammonium Iodide (PEAI) was purchased from Xi’an Polymer Light Technology Corporation.  \textit{n}-butylammonium iodide (\textit{n}-BAI) and \textit{n}-octylammonium Iodide (\textit{n}-OAI) were purchased from Dysol (Australia).

\textit{2D Perovskite Thin-Film Fabrication}: Quartz substrates were cleaned using sonication and rinsing in distilled water, acetone, and isopropanol. Following this, the substrates were treated by UV-Ozone for 15 mins. The optimised 1M \ce{(BA)2PbI4}, \ce{(OA)2PbI4}, and \ce{(PEA)2PbI4} 2D perovskite precursor solutions were prepared by dissolving the mixture of 2:1 (PEAI:\ce{PbI2},\textit{n}-BAI:\ce{PbI2}, and \textit{n}-OAI:\ce{PbI2}) in DMF:DMSO=4:1 (v:v). The compact 2D perovskite layers were deposited on the substrates by spin-coating at 5000 rpm for 30 s, with an acceleration of 4000 rpm/s. Following this, the coated substrates were transferred onto a hotplate and annealed at 100 \degree C for 5 mins.

\textit{Photoluminescence Spectroscopy}: The photoluminescence was measured using a 405 nm Thorlabs M405L2 mounted LED for the excitation beam. The excitation was reflected from a 400 nm dichroic mirror, and the collection is obtained through the same lens as the excitation. The collection is then passed through a 450 nm long-pass filter and into a Thorlabs CS200 fibre-coupled spectrometer. The temperature was controlled using a cryogen-free variable temperature cryostat (Cryogenic Ltd), with a model 350 temperature controller (Lakeshore). The magnetic field was applied in the Voigt configuration using an electromagnet as part of a E500 electron spin resonance system (Bruker). 

\textit{UV-Vis Spectrophotometry}: UV-visible absorbance of the thin films was measured using UV-2600 Shimadzu spectrophotometer with integrating sphere attachment ISR2600. 

\textit{X-ray Diffraction}: X-Ray diffraction patterns were measured by a PANalytical Xpert materials research diffractometer (MRD) with Cu K$\alpha$ radiation ($\lambda$ = 1.54056 Å), using an accelerating voltage of 45 kV and a current of 40 mA.

\begin{acknowledgement}

D.R.M., C.G.B., L.V.G., and N.S. acknowledge the support of the Australian Research Council (ARC) Centre of Excellence in Exciton Science (CE170100026). X.H. acknowledges financial support by the ARC Future Fellowship (FT190100756). A.M.S. acknowledges the funding support from ACAP (RG193402-I). The authors acknowledge use of facilities in the Solid State and Elemental Analysis Unit at Mark Wainwright Analytical Centre.

\end{acknowledgement}

% \begin{suppinfo}
% \end{suppinfo}

\bibliography{references}

\end{document}

\renewcommand{\thefigure}{S\arabic{figure}} %Changes the figure labelling 
\setcounter{figure}{0} %Sets figure counter to zero

\begin{figure*}[htbp]
  \centering
  \includegraphics[width=0.95\textwidth]{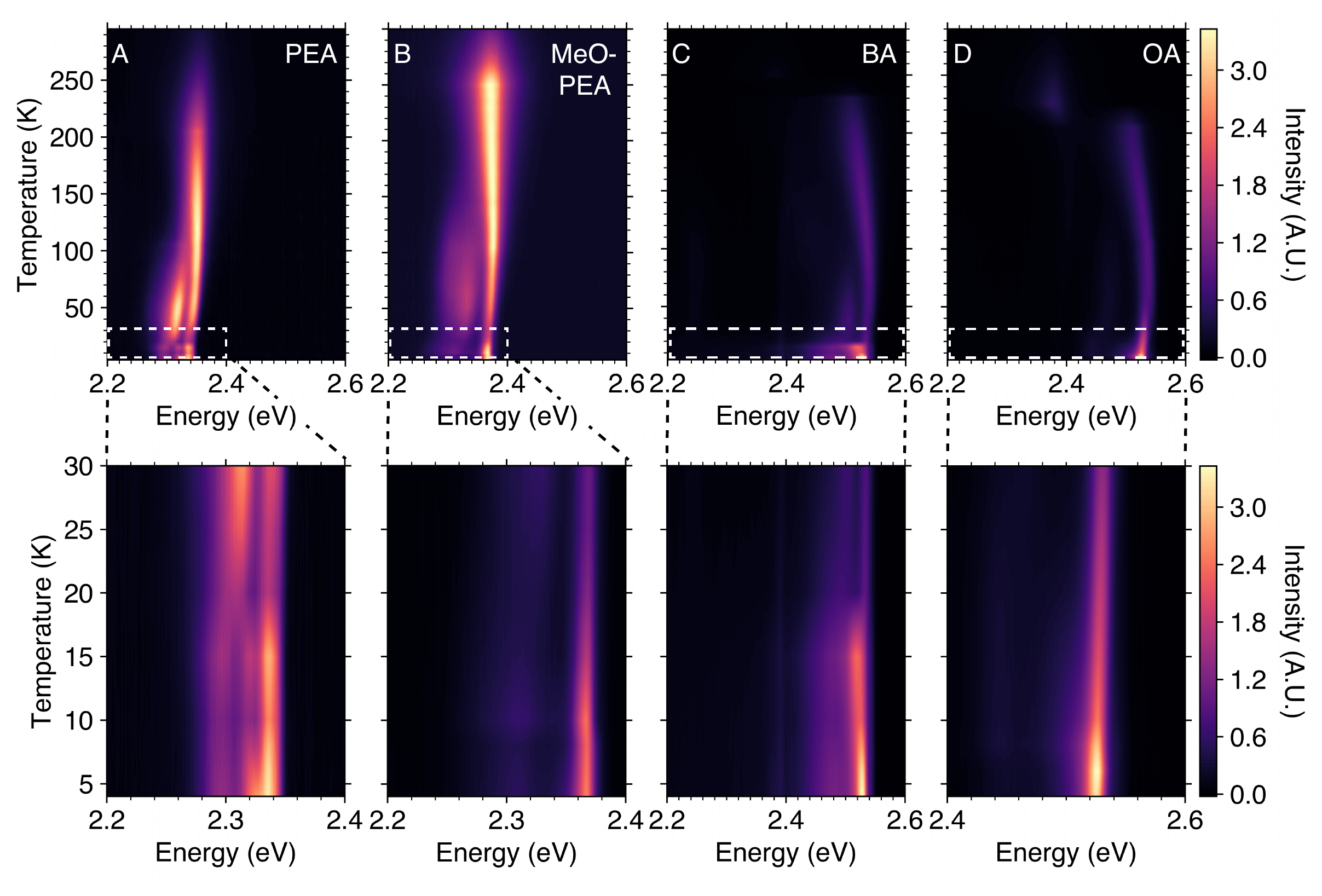}
  \caption{\textbf{Temperature-dependent photoluminescence.} False-colour map of photoluminescence spectra for (A) \ce{(PEA)2PbI4}, (B) \ce{(MeO-PEA)2PbI4} (C) \ce{(BA)2PbI4}, and (D) \ce{(OA)2PbI4}. Cutouts on bottom show enlarged maps of spectra up to a temperature of 30 K.}
  \label{fig:tempdeps_nonnorm}
\end{figure*}

\begin{figure*}[htbp]
  \centering
  \includegraphics[width=0.75\textwidth]{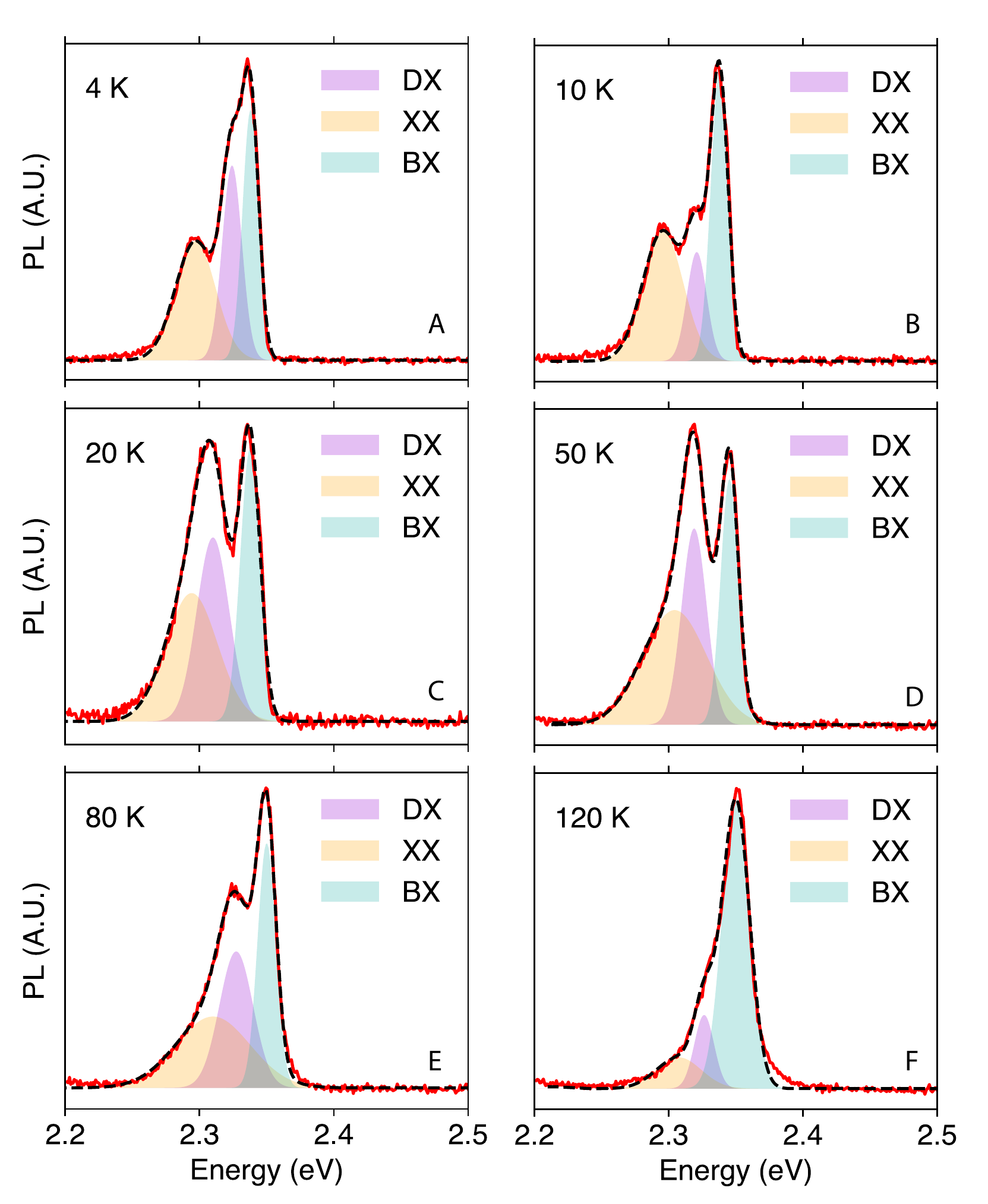}
  \caption{\textbf{Modelling of temperature-dependent photoluminescence of \ce{(PEA)2PbI4}.} Temperature-dependent photoluminescence spectra (red) fitted with the sum of three Gaussian contributions (black) from the dark excitons (DX), biexcitons (XX), and bright excitons (BX) for A) 4K, B) 10 K, C) 20 K, D) 50 K, E) 80 K, and F) 120 K.}
  \label{fig:tempslice_PEPI_PL}
\end{figure*}

\begin{figure*}[htbp]
  \centering
  \includegraphics[width=0.75\textwidth]{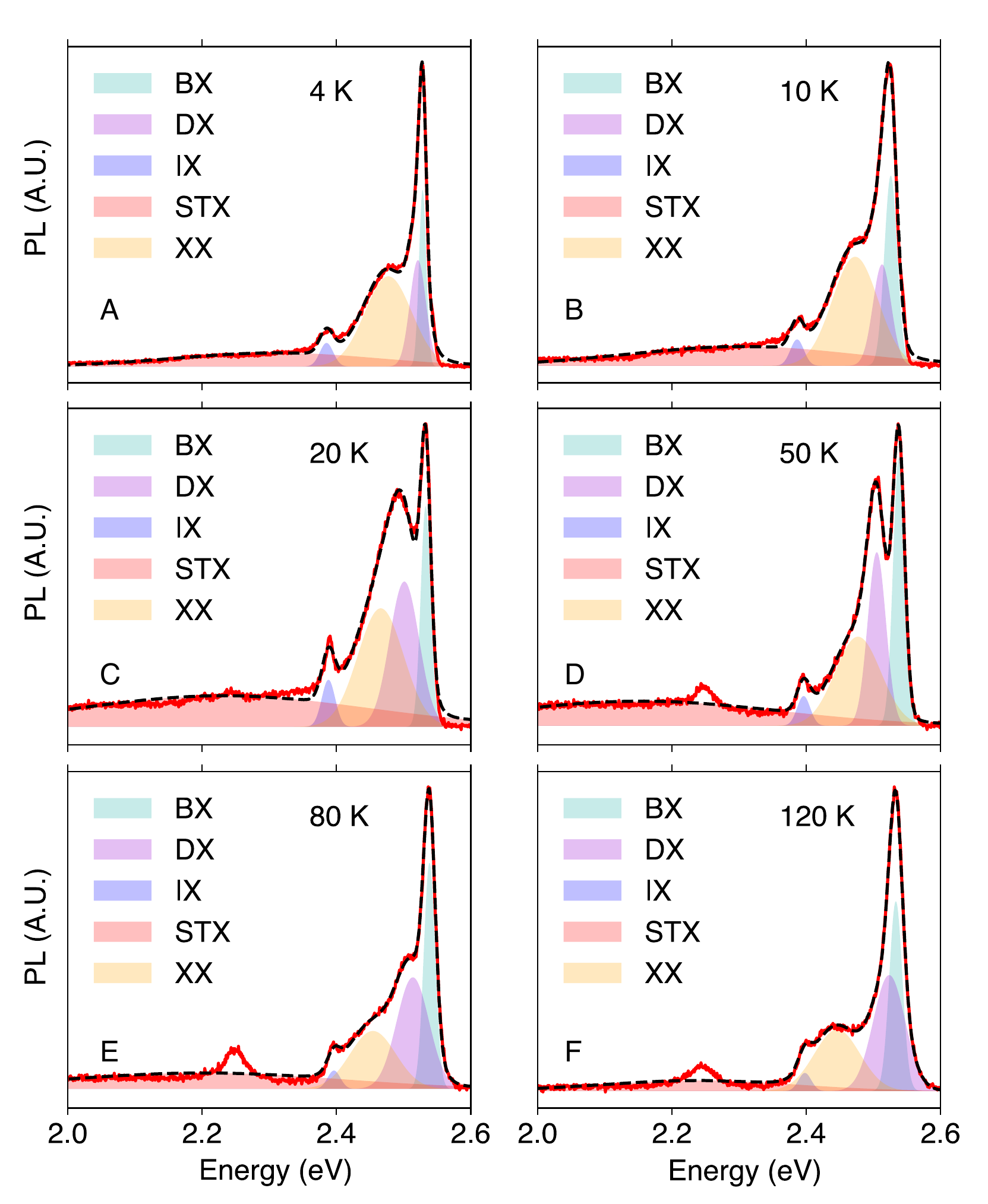}
  \caption{\textbf{Modelling of temperature-dependent photoluminescence of \ce{(BA)2PbI4}.} Temperature-dependent photoluminescence spectra (red) fitted with the sum of three Gaussian contributions (black) from the dark excitons (DX), biexcitons (XX), bright excitons (BX), interior excitons (IX), self-trapped/defect excitons (STX) for A) 4K, B) 10 K, C) 20 K, D) 50 K, E) 80 K, and F) 120 K.}
  \label{fig:tempslice_BA_PL}
\end{figure*}

\begin{figure*}[htbp]
  \centering
  \includegraphics[width=0.75\textwidth]{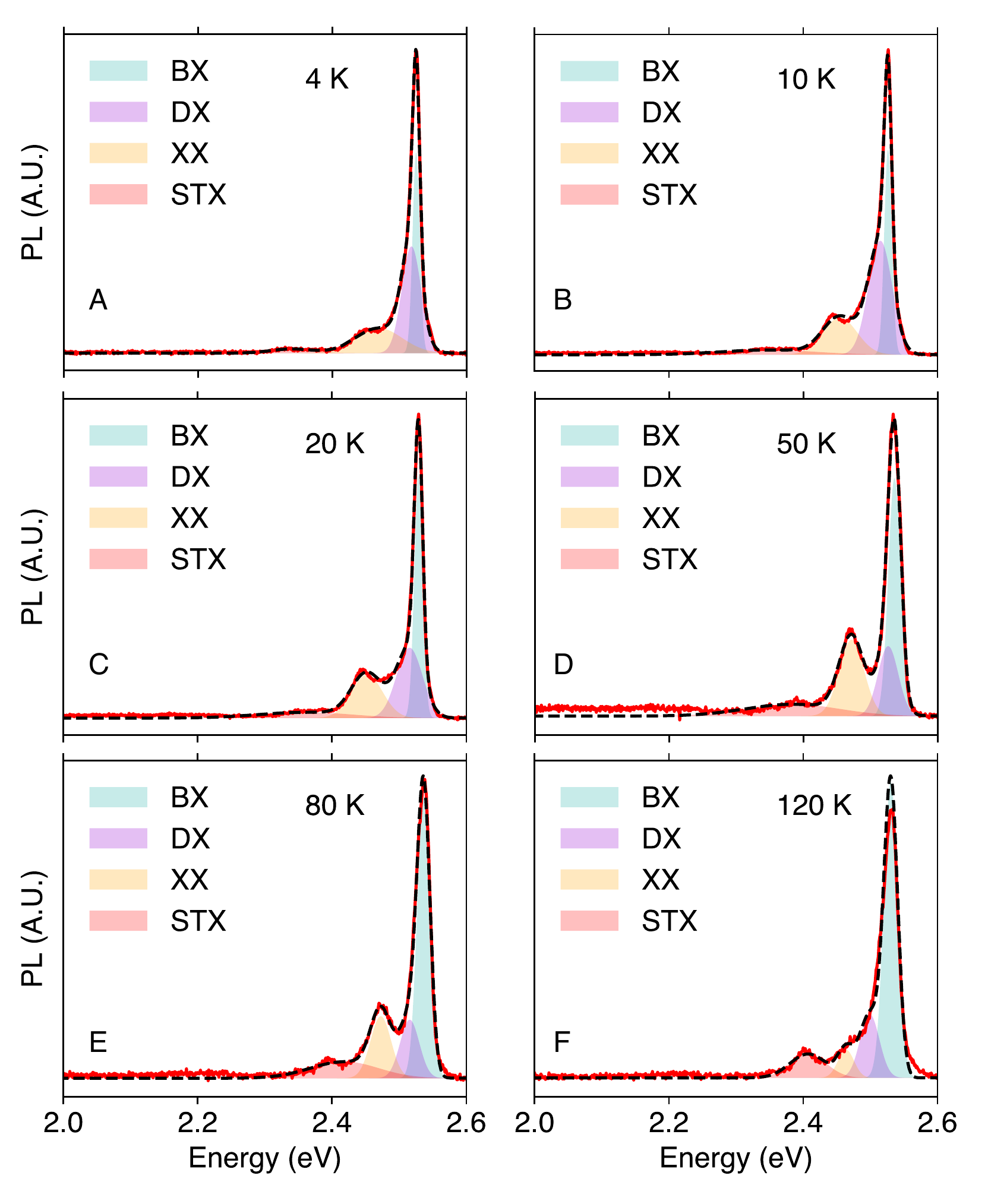}
  \caption{\textbf{Modelling of temperature-dependent photoluminescence of \ce{(OA)2PbI4}.} Temperature-dependent photoluminescence spectra (red) fitted with the sum of three Gaussian contributions (black) from the dark excitons (DX), biexcitons (XX), bright excitons (BX), self-trapped/defect excitons (STX) for A) 4K, B) 10 K, C) 20 K, D) 50 K, E) 80 K, F) and 120 K.}
  \label{fig:tempslice_OA_PL}
\end{figure*}

\begin{figure*}[htbp]
  \centering
  \includegraphics[width=0.75\textwidth]{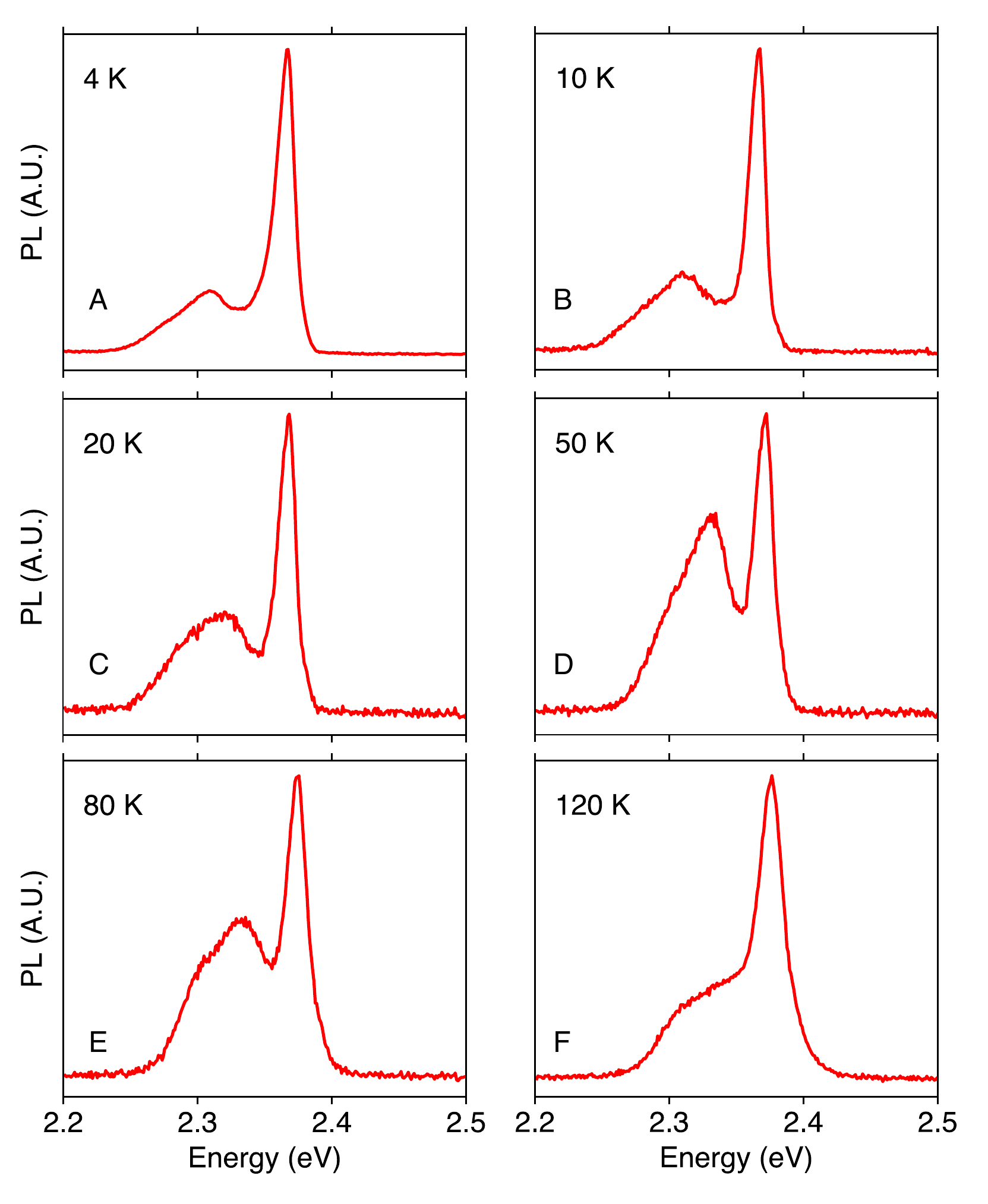}
  \caption{\textbf{Temperature-dependent photoluminescence of \ce{(MeO-PEA)2PbI4}.} Temperature-dependent photoluminescence spectra for A) 4K, B) 10 K, C) 20 K, D) 50 K, E) 80 K, and F) 120 K.}
  \label{fig:tempslice_MeOPEA_PL}
\end{figure*}

\begin{figure*}[htbp]
  \centering
  \includegraphics[width=0.75\textwidth]{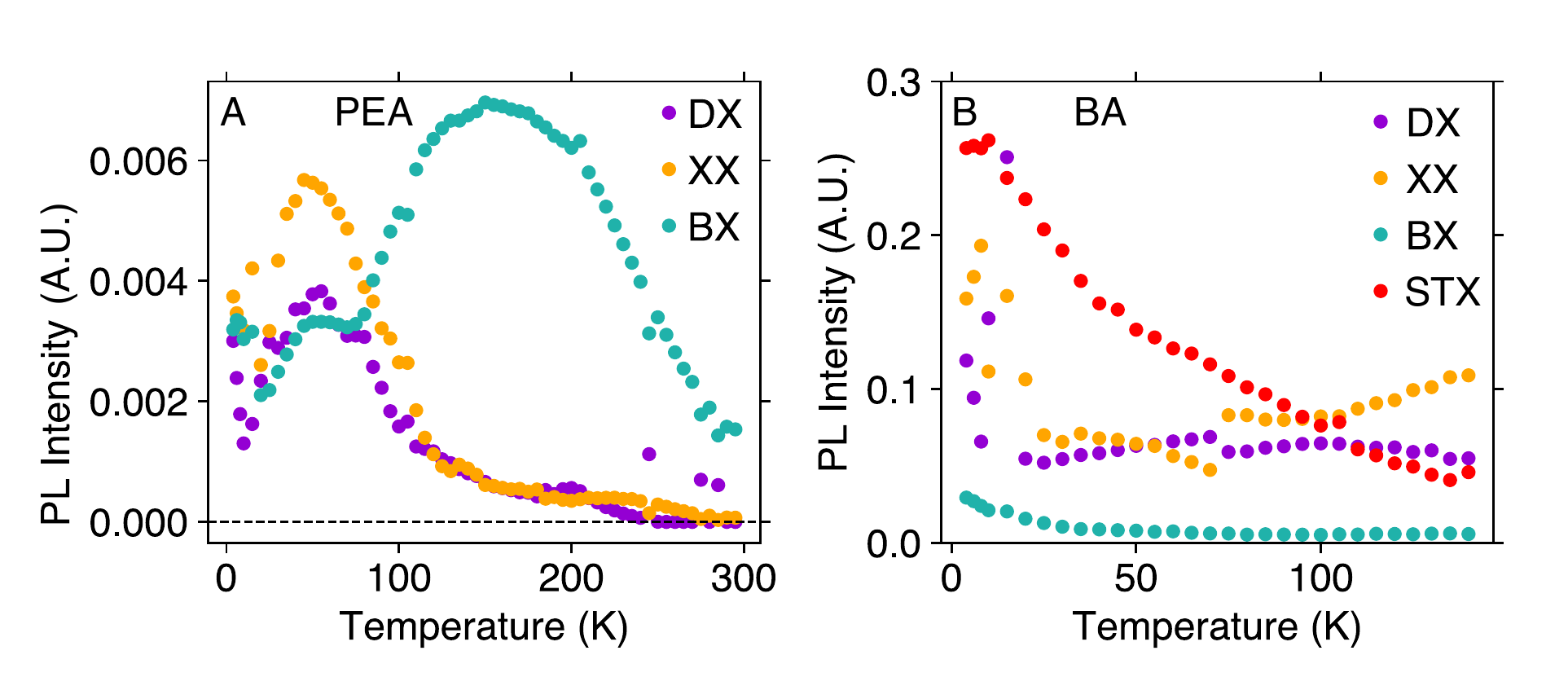}
  \caption{\textbf{Temperature-dependent photoluminescence intensity from fitting results} Integrated intensity for individual transitions representing the bright exciton (BX), dark exciton (DX), biexciton (XX), and self-trapped/defect exciton (STX) states as a function of temperature for (A) \ce{(PEA)2PbI4} and (B) \ce{(BA)2PbI4}} 
  \label{fig:intgint_tempdep}
\end{figure*}

\begin{figure*}[htbp]
  \centering
  \includegraphics[width=0.75\textwidth]{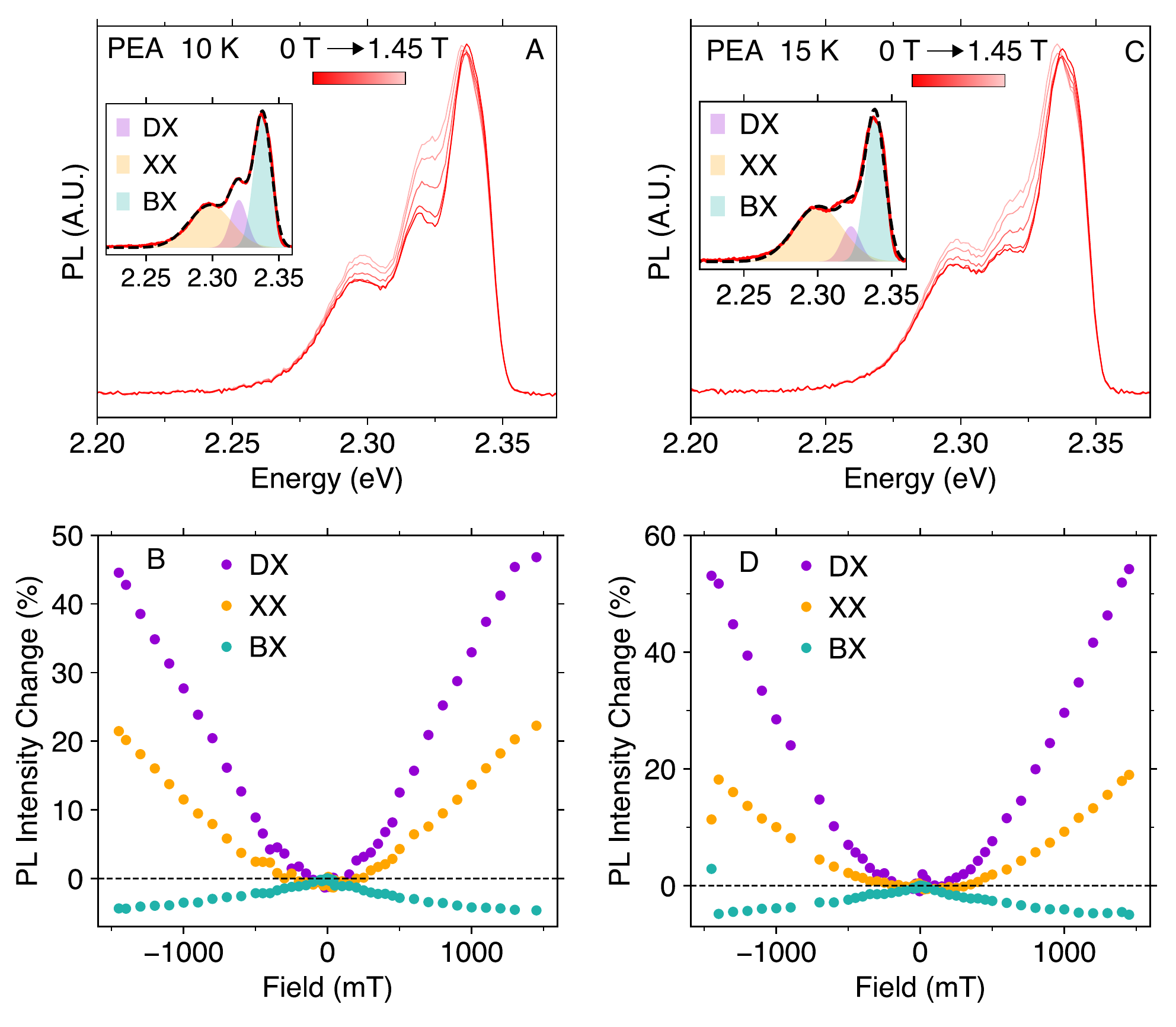}
  \caption{\textbf{Response of individual exciton transitions to an applied in-plane magnetic field.} Magnetic-field dependent photoluminescence spectra of \ce{(PEA)2PbI4} at (A) 10 K and (C) 15 K. Integrated intensity (amplitude) of Gaussian components versus magnetic field for the individual bright exciton (BX), dark exciton (DX), and biexciton (XX) transitions in \ce{(PEA)2PbI4} at (B) 10 K and (D) 15 K.}
  \label{fig:MPL_10kand15K}
\end{figure*}

\begin{figure*}[htbp]
  \centering
  \includegraphics[width=0.6\textwidth]{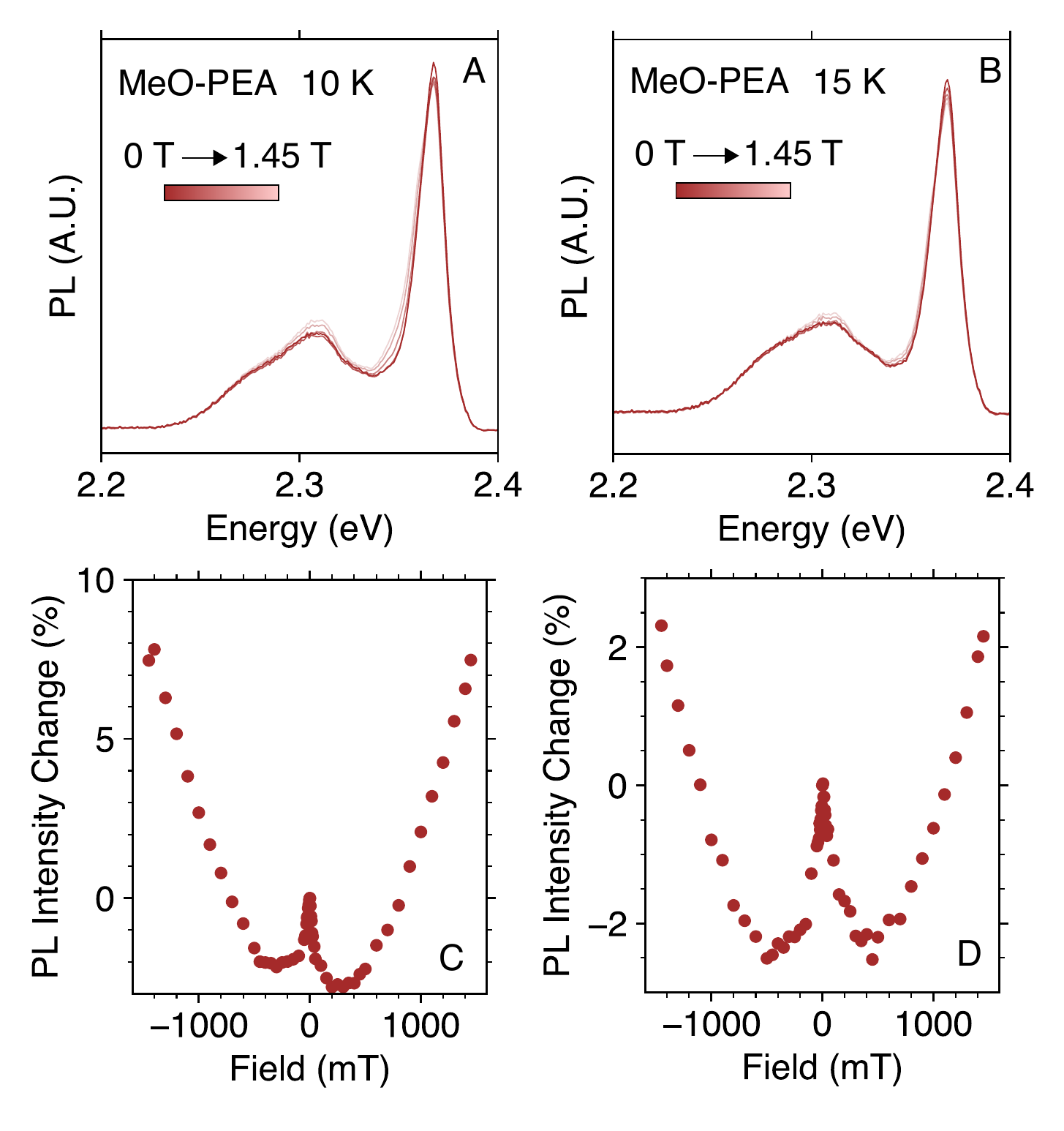}
  \caption{\textbf{Photoluminescence spectra in an applied magnetic field for different temperatures.} Magnetic-field dependent photoluminescence spectra of \ce{(MeO-PEA)2PbI4} at (A) 10 K and (C) 15 K. Integrated intensity of photoluminescence versus magnetic field at (B) 10 K and (D) 15 K.}
  \label{fig:MEOMPL_10kand15K}
\end{figure*}

\begin{figure*}[htbp]
  \centering
  \includegraphics[width=0.75\textwidth]{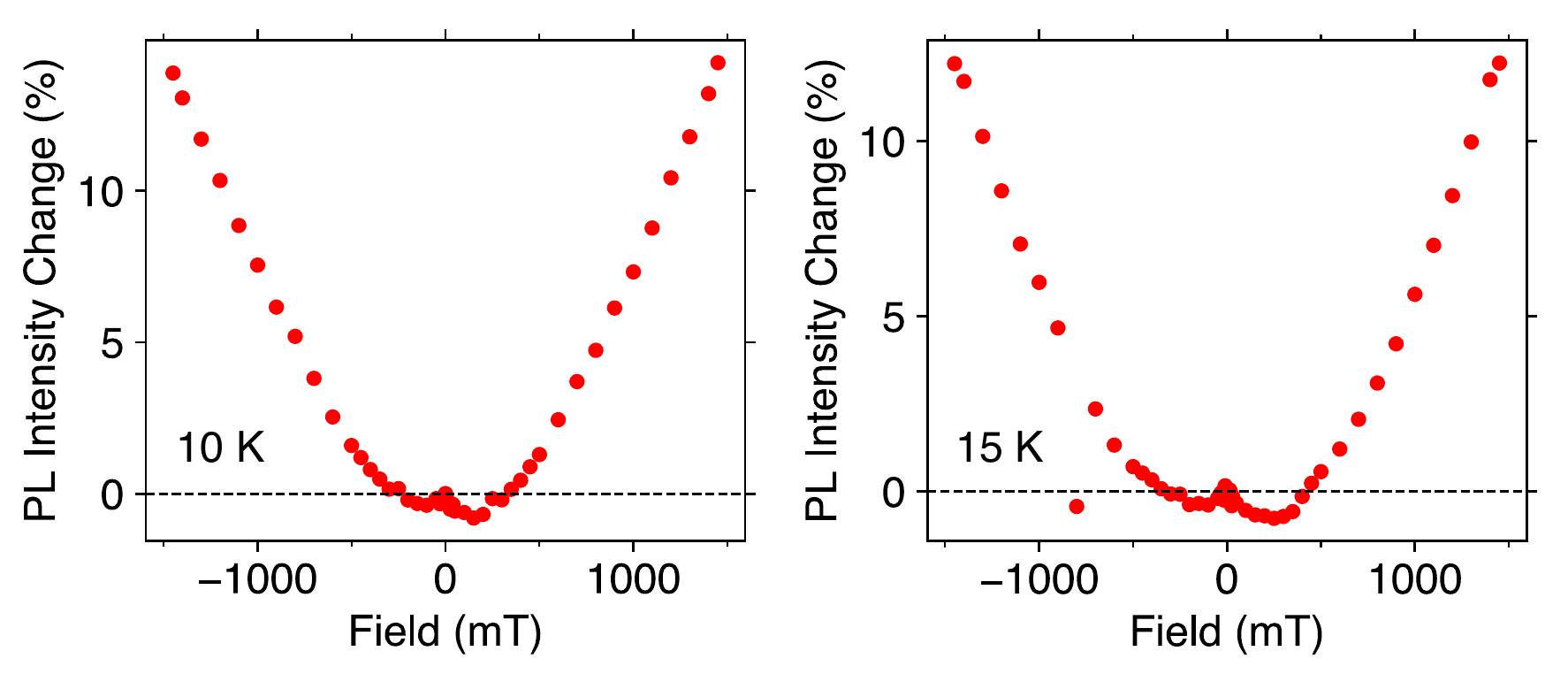}
  \caption{\textbf{Total integrated photoluminescence versus applied magnetic field for \ce{(PEA)2PbI4} at (A) 10 K and (B) 15 K.}}
  \label{fig:TotalMPL_15K}
\end{figure*}
\clearpage

\subsection{Fitting of photoluminescence spectra}

The integrated intensity change is calculated by first fitting each measured spectra with the sum of multiple Gaussian functions, which are attributed to the bright, dark, and biexciton emission:  

\begin{equation}\label{eq:3G}
    I(E)=\sum_{}{}A_{i}e^{-\frac{(E-E_{i})^2}{2\sigma_i^2}}, \tag{S1}
\end{equation}

 \noindent where $A_i$ is the amplitude, $E_i$ is the peak centre energy, and $\sigma_i$ is the standard deviation. $i$ represents the individual exciton transitions such as bright exciton, dark exciton, and biexciton states. The amplitude of these individual Gaussian functions is used to calculate the percentage change in photoluminescence ($\frac{\Delta PL}{PL}$) as a function of an external magnetic field using the expression,

\begin{equation}\label{eq:ampmpl}
    \frac{\Delta PL}{PL}= \frac{A(B)-A(0)}{A(0)}\times 100 \%,
    \tag{S2}
\end{equation}

 \noindent where $A(B)$ is the amplitude at a given magnetic field $B$, and $A(0)$ is the zero-field amplitude for a given exciton transition.

\subsection{Determining the biexciton binding energy}

The decay rate constant ($k_{ex}$) for excitons can be expressed as the sum of individual contributions, given by,

\begin{equation}\label{eq:kex}
    k_{ex} = k_r + k_{nr} + k_{dis}e^{\left(\frac{-E_B}{k_BT}\right)},
    \tag{S3}
\end{equation}

 \noindent where $k_r$ and $k_{nr}$ are the temperature-independent radiative and nonradiative decay rate constants, respectively, $k_{dis}$ is the temperature-dependent dissociation constant, and $k_B$ is the Boltzmann constant. $E_B$ is the 'activation energy' for exciton dissociation, which is equivalent to the exciton binding energy. Given that the photoluminescence quantum yield $\phi$ at a certain temperature $T$ and 0K can be defined by $k_r/k_{ex}$ and $k_r/k_r + k_{nr}$, respectively, the inverse of PL emission quantum yield, $1/\phi(T)$, as a function of temperature thus can be expressed as:

 \begin{equation}\label{eq:phi}
 1/\phi(T) = 1/\phi(T = 0) + (k_{dis}/k_r)e^{\left(\frac{-E_B}{k_BT}\right)} 
 \tag{S4}
\end{equation}
 
 The biexciton emission component of the PL is extracted from equation \ref{eq:kex} and its temperature-dependent integrated intensity is determined. The biexciton binding energy $E_B$ can be determined by fitting the following equation:

 \begin{equation}
  1 / I(T) = A + Be^{\left(\frac{-E_B}{k_BT}\right)},
  \tag{S5}
 \end{equation}
 
 \noindent where A corresponds to the inverse PL intensity $(1 / I_0)$ at $T = 0 K$, and B is equal to $k_{dis}/k_{r}$. 

\begin{table}
  \caption{Fitting parameters for exciton binding energy ($E_B$) of the biexciton emission in \ce{(PEA)2PbI4}}
  \label{tbl:example}
  \begin{tabular}{ll}
    \hline
    Fitting parameters  & Obtained values ± standard error  \\
    \hline
    $E_B$   & 40.18 ± 3.24 meV   \\
    A $(1/I_0)$ & 161 ± 52.86  \\
    B $(k_{dis}/k_r)$  & 27291 ± 5395  \\
    \hline
  \end{tabular}
\end{table}